# STW and SPIHT Wavelet compression using MATLAB wavelet Tool for Color Image

M. Tiwari, Asst. Professor, *St. Aloysius College, Jabalpur*

*Abstract*— Images can be represented by mathematical function using wavelets. Wavelet can be manipulated (shrink/expand) by applying some values to its function. It helps to localize the signals. Application of wavelet in images processing has larger scope as proved. Image compression is one of the dimension. There are various wavelet image compression techniques. This research paper focused on comparison of only two techniques i.e. STW and SPIHT for color JPEG images.

## I. INTRODUCTION

Digital world using more visual/image for securities, geospatial, education, entertainment and in many more areas. Images are quite expensive to maintain as compare than textual data. The new challenges are at door steps to maintain these images, while users are having limited resources. Maintenance has two dimension first keeping images into permanent storage devices as records and second is transporting the images between various geographical locations using different type of computer network or storage devices.

As quality of image are increased requirement of resources are also increased in same ratio. To make maintenance cheap, images quality can't be downgraded.

To save time and resources, different efficient image compression techniques are used to reduce the size of the still images. Compression reduces the image size for better resource management.

This research paper focuses on comparison of two wavelet based image compression techniques which are SPIHT and STW. Experiments are done on JPEG format color image. Obtained results are arranged into matrices with image compression quality measure i.e. MSE, PSNR, CR and size.

SPIHT algorithm is a one of the version of the EZW algorithm, It was introduced by Said and Pearlman in 1996. SPIHT stands for set partitioning in hierarchical trees. Hierarchical trees means quad trees that is defined in EZW discussion. Set partitioning refers to the way quad trees partitions the wavelet transform values at a given limit values. [1]. Objective was to produce better compression result for still images.

The discussion of SPIHT is consist of three parts. First, describes a modified version of the algorithm introduced in Said and Pearlman [44] and it was referred as the spatial-orientation tree wavelet (STW) algorithm. STW is essentially the SPIHT algorithm; the only difference is that SPIHT is slightly more careful in its organization of coding output. Second, SPIHT algorithm is describes that it is easier to explain SPIHT using the concepts underlying STW.

Third is how well SPIHT compresses images. The only difference between STW and EZW is that STW uses a different approach to encoding the zero tree information. STW uses a state transition model. From one value to the next, the locations of transform values undergo state transitions, this model allows STW to reduce the number of bits needed for encoding [1].

## II. LITERATURE REVIEW

T. Kumar and D. Choudhary experimented on six wavelet based compression techniques: ezw, spiht, stw, wdr, aswdr and spiht_3d found that the spiht_3d technique performs better than other wavelet techniques. They opt a color image. [7]

P. A. Babu and Dr. K.V.S.V.R. Prasad, used three algorithms for image compression JPEG, SPIHT and Modified BPT for two parameter i.e. PSNR and CR. Modified BPT is producing better results for PSNR and CR values. The paper provides the proposed methodology for the compression of image to be used more effectively which is capable of providing much efficient quality metrics values and visual quality as compared to the existing expression techniques JPEG and SPIHT. [8]

R. Kumar and Dr. S. Singh has concluded his paper WDR gives better result than STW for image compression [9].

## III. PRELIMINARIES

*A. Wavelet*

Wavelet may be seen as a complement to classical Fourier decomposition method [5]. Suppose a certain class of function is given and we want to find simple function $f_0, f_1, f_2, f_3\ldots$ such that

$$f(x) = \sum_{\infty}^{\infty} a_n f_n(x) \qquad (1)$$
$$a_n = \text{coefficient}$$

Wavelet is a mathematical tool leading to representation of the type (1) for a large class of functions $f$.

Wavelet means small wave (the sinusoidal used in Fourier analysis are big wave) and a wavelet is an oscillation that decays quickly.

$$\int_{\infty}^{\infty} |\psi(t)|^2 \, dt < \infty \quad (2)$$

$$\int_{\infty}^{\infty} |\psi(t)| \, dt = \infty \quad (3)$$

[5].

*B. SPIHT*

This is a highly refined version of the EZW algorithm. It was introduced by Said and Pearlman in 1996. SPIHT stands for set partitioning in hierarchical trees. Set partitioning refers to the way these quad trees partition the wavelet transform values at a given threshold. [1].

*C. STW*

It is another alternative for wavelet based image compression technique. It is essentially the SPIHT algorithm; the only difference is that SPIHT is slightly more careful in its organization of coding output. It uses a state transition model. [1].

*D. PSNR*

Stand for Peak Signal to Noise Ratio. This is one of the most popular and commonly used measurements of reconstruction of lossy image compression. The signal is original data and noise is the error introduced by compression.

The ratio is often used as a quality measurement between original image and compressed image. The higher PNSR better the quality of the compressed or reconstructed image [3]

*E. MSE*

Stands for Mean Square Error. This represents the cumulative squared error between the compressed and original image. The lower the value of MSE, the lower the error [4].

*F. CR*

Stand for Compression Ratio. It is ratio of non-zero element of original matrices and transformed matrix. Every image is a representation of bits. These bits are arranged in the matrix form. The bits were used to represent original and compressed image are compared.

Compression Ratio = Original Image/Compressed Image;

*G. MATLAB*

MATLAB is an interactive software whose basic data element is an array that does not require dimensioning. This helps to resolve many technical computing related problems, specifically concerned with matrix and vector formulations, in a fraction of the time it would take to write a program in a scalar non interactive language [2].

*H. Wavelet Toolbox*

Wavelet Toolbox provides functions and apps for analyzing and synthesizing signals and images. The toolbox includes algorithms for continuous wavelet analysis, wavelet coherence, synchro squeezing, and data-adaptive time-frequency analysis [6]. The toolbox contains applications and functions for wavelet analysis of signals and images, also covers to wavelet packets and dual-tree transforms [6].

## IV. RESULT ANALYSIS

Four parameter MSE, PSNR, CR and Size is used to evaluate the quality of the compressed image by MATLAB wavelet tool box. The 256X256 goddess image was selected for experiments.

Compressed image results are recorded for different eight level. TABLE-I and TABLE-II demonstrate the result of experiment obtained. Overall result discussion is following for all the parameter.

Fig 1. Image size 256X256 (Pixel) and occupies space 18.4 KB in memory.

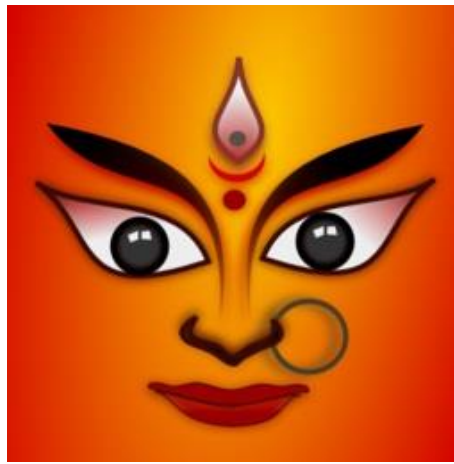

**TABLE-I**
IMAGE COMPRESSION RESULT SPIHT

| Levels | MSE | PSNR | CR | Size (KB) |
|---|---|---|---|---|
| 1 | 4.387 | 41.71 | 77.94 | 9 |
| 2 | 7.445 | 39.41 | 26.59 | 9 |
| 3 | 16.98 | 35.83 | 10.98 | 8 |
| 4 | 38.62 | 32.26 | 5.26 | 8 |
| 5 | 96.5 | 28.29 | 2.56 | 8 |
| 6 | 223.8 | 24.63 | 1.16 | 7 |
| 7 | 449.5 | 21.6 | 0.53 | 5 |
| 8 | 868.6 | 18.74 | 0.21 | 3 |

contains the result obtained by applying the SPIHT compression algorithm on this color image. The operation is applied at eight different levels.

1) STW is producing better compression data for PNSR than SPIHT.
2) SPIHT, MSE value is 4.387 and STW shows 0.9114 MSE means STW handling compression in better ways.
3) Up to third level SPIHT is show comparatively better than software but for afterwards STW is doing better.
4) Size of compression image is same for better techniques only different at level three.

Image is compressed and results are recorded for different eight levels. STW is producing far better PSNR value then SPIHT for every level of compression. SPIHT producing the 4.387 while STW is resulting only 0.9114 values for MSE. CR values for STW is much better than SPIHT. TABLE-I and TABLE-II contains the result obtained at compression.

**TABLE II**
**IMAGE COMPRESSION RESULT STW**

| Levels | MSE | PSNR | CR | Size(KB) |
|---|---|---|---|---|
| 1 | 0.9114 | 48.53 | 54.34 | 9 |
| 2 | 3.35 | 42.88 | 24.15 | 9 |
| 3 | 9.983 | 38.14 | 12.16 | 9 |
| 4 | 27.64 | 33.72 | 6.44 | 8 |
| 5 | 76.21 | 29.31 | 3.30 | 8 |
| 6 | 191.5 | 25.31 | 1.55 | 7 |
| 7 | 401.3 | 22.1 | 0.69 | 5 |
| 8 | 806.2 | 19.07 | 0.28 | 3 |

contains the result obtained by applying the STW compression algorithm on this color image. The operation is applied at eight different levels.

## V. CONCLUSION

Average of MSE, PSNR and CR values for STW is 189.63, 12.86 and 32.35 and for SPIHT is 213.229, 30.30 and 15.65. Obtained results show that STW is better wavelet image compression technique than SPIHT for color image.

## VI. ACKNOWLEDGMENT

I pay my sincere thanks to my guide and mentor Professor S. S. Pandey, Assistant Professor, RDVV, MP, India.

I am thank full to Rev. Dr. G. Vazhan Arasu, Principal and Mrs. Siby Samuel, HOD, DCSA, St. Aloysius College, Jabalpur for their persistent motivation and inspiration.

I am also thank full to Dr. J.W. Shareef and Dr. Ashish Parashar, Technical Faculty, Department of Mathematics, Rani Durgwati University, Jabalpur, MP and all of my colleagues for their consistently support.

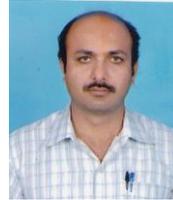

**Manish Tiwari** was born in Jabalpur, Madhya Pradesh, Bharat in 1983. He received the Bachelor in Computer Application from Maharishi Mahesh Yogi Vedic University in 2003 and Masters in Computer Application degree in 2006 from RTM, Nagpur University and M. Phil in Computer Science in 2014 from Rani Durgawati University, Jabalpur.

He worked as Software Engineer for more six year in various companies like Mahindra Satyam, Venture Infotek Private Limited, Virtal Galaxy, CMC limited. Presently, working as Assistant Professor in St. Aloysius' College (Auto), Jabalpur, MP, Bharat since July-2014. He has published two research papers.